\documentclass[%
aip,
jcp,%
amsmath,amssymb,
reprint,%
]{revtex4-1}
\usepackage{color}

\newcommand{\sub}[1]{_{\mbox{\scriptsize {#1}}}}

\usepackage{graphicx}% Include figure files
\usepackage{dcolumn}% Align table columns on decimal point
\usepackage{bm}% bold math
\def\siml{\hspace{1ex} ^{<} \hspace{-2.5mm}_{\sim} \hspace{1ex}}

\begin{document}

%\preprint{AIP/123-QED}
\preprint{}

\title[Nucleation of  water by MD simulations]{Molecular Dynamics Simulations of the Nucleation of Water: 
Determining the Sticking Probability and Formation Energy of a Cluster}% Force line breaks with \\

\author{Kyoko K. Tanaka}
\email{kktanaka@lowtem.hokudai.ac.jp}%Lines break automatically or can be forced with \\
%\homepage{}
\affiliation{Institute of Low Temperature Science, Hokkaido University, Sapporo 060-0819, Japan}

\author{Akio Kawano}
\affiliation{Japan Agency for Marine-Earth Science and Technology, Kanagawa, Japan}

\author{Hidekazu Tanaka}
\affiliation{Institute of Low Temperature Science, Hokkaido University, Sapporo 060-0819, Japan}

\date{\today}% It is always \today, today,

\begin{abstract}
We performed molecular dynamics (MD) simulations of the nucleation of
water vapor in order to test nucleation theories.  Simulations were
performed for a wide range of supersaturation ratios ($S = 3$-$25$)
and water temperatures ($T\sub{w}=300$-$390$~K).  We obtained the
nucleation rates and the formation free energies of a subcritical
cluster from the cluster size distribution.  The classical nucleation
theory (CNT) and the modified classical nucleation theory (MCNT)
overestimate the nucleation rates in all cases.  The
semi-phenomenological (SP) model, which corrects the MCNT prediction
using the second virial coefficient of a vapor, reproduces the
formation free energy of a cluster with the size $\siml 20$ to within
10 \% and the nucleation rate and cluster size distributions to within
one order of magnitude.  The sticking probability of the vapor
molecules to the clusters was also determined from the growth rates of
the clusters.  The sticking probability rapidly increases with the
supersaturation ratio $S$, which is similar to the Lennard-Jones
system.
\end{abstract}

%\pacs{05.10.-a, 05.70.Fh, 05.70.Ln, 05.70.Np, 36.40.Ei, 64.60.qe, 64.70.Hz, 64.60.Kw, 64.10.+h, 83.10.Mj, 83.10.Rs, 83.10.Tv}% PACS, the Physics and Astronomy
\keywords{water, molecular dynamics method, nano-clusters, nucleation, phase transitions, liquid-vapour transformations, sticking probability}%Use showkeys class option if keyword
\maketitle

\section{INTRODUCTION}

The classical nucleation theory (CNT) is the most widely used model
for describing homogeneous nucleation and provides the nucleation rate
as a function of the supersaturation ratio and the surface tension of
a condensed phase\cite{volmer,beck,zeld,feder}. 
 However, many studies have reported that
the CNT fails to describe experimentally obtained 
results\cite{schmitt1,schmitt2,miller,adams,dill,oxtoby,delale,viisanen1993,wright,laaksonen,viisanen1994,kane,kaarle,luijten,anisimov2001,wolk,mikheev,khan,kim,holten,brus2008,brus2009,manka}. 
In the case of water, the deviations between the nucleation rates
predicted by the classical nucleation theory and the experimental
values are in the order of $10^2$-$10^3
$\cite{viisanen1993,miller,luijten,wolk,mikheev,khan,kim,holten,brus2008,brus2009,manka}.  The nucleation rates obtained
by molecular dynamics (MD) and Monte Carlo (MC) simulations also
significantly differ from the predictions of the CNT\cite{ford1997,kusaka1998,ym1,ym2,wolde1998,kusaka1999,oh1999,senger,wolde1999,laasonen,oh2000,tanimura,vehkam,chen2001,schaaf,yoo,toxvaerd,toxvaerd2,tanaka2005,matsubara2007,tanaka2011,diemand2013}.  The
nucleation rate is governed by the formation free energy of a critical
cluster, which is the smallest thermodynamically stable cluster.  In
the classical nucleation theory, the formation free energy is simply
evaluated using the surface energy of the bulk material.  Since the
critical clusters are considered to be nano-sized, the error in the
classical theory is thought to come from the difference in the
properties of such a small cluster and a bulk material.

In previous studies, there have been significant advances in
theoretical models of homogeneous nucleation\cite{dill, oxtoby,
 laaksonen, 
 delale,laasonen,reguera,reguera-reiss,kalikmanov2006,
wedekind2007,merikanto2007,kalikmanov2008}.  One
of the most successful and useful models is the semi-phenomenological
(SP) model\cite{dill,delale,laaksonen},
 which corrects the evaluation of the formation
energy of a cluster in the CNT by using the second virial coefficient
of a vapor. The predictions of the SP model agree well with
experimental data for various substances including
water\cite{dill,delale,laaksonen}.  By performing MD simulations of nucleation
for Lennard-Jones systems, Tanaka et al.\cite{tanaka2005,tanaka2011} 
tested the SP model
and found that it also agrees well with their MD simulations. However,
only large supersaturation ratios were possible in their MD
simulations.  Recent large scale simulations by Diemand et
al.\cite{diemand2013} found some deviations in the nucleation rate
from the SP model in the case of a small supersaturation ratio.

Besides Lennard-Jones systems, many other MD simulations examining the
nucleation process of water molecules have been carried out.  Yasuoka
and Matsumoto\cite{ym2} investigated homogeneous nucleation for the
first time using MD simulations of water molecules.  The nucleation
rate they obtained at 350 K was two orders of magnitude less than that
predicted by the classical theory.  Matsubara et
al.\cite{matsubara2007} carried out MD simulations at various
temperatures and supersaturation ratios, using a simple point
charge/extended (SPC/E) water model\cite{berendsen}.  They measured
the nucleation rate, the critical nucleus size, and the formation free
energy of a cluster and compared them with the various theoretical
models.  They showed that all theoretical models (CNT, SP model, and
the scaled model) predict the nucleation rates obtained by the
simulations to within one or two orders of magnitude.  On the other
hand, the formation free energy of a cluster derived by the
simulations was considerably larger than that obtained with the
theoretical models.  They showed that the deviations in the formation
free energy can be explained by the larger growth rate of clusters
than that assumed by the theories.  Recently, Zipoli et
al.\cite{zipoli} developed a new coarse-grained model for water to
study nucleation from the vapor and obtained smaller nucleation rates
than the previous studies.

In addition to the formation free energy of a cluster, the sticking
probability $\alpha$ of vapor molecules to clusters is another
important factor for determining the nucleation rate.  The sticking
probability is usually assumed to be unity, despite the fact that the
nucleation rate is proportional to $\alpha$ in the nucleation theory.
In the previous studies\cite{tanaka2011,diemand2013}, 
the sticking probability in a Lennard-Jones
type system was examined through MD simulations, by observing the
growth rate of stable clusters larger than the critical size.  They
showed that the sticking probability decreases with decreasing
supersaturation ratio; however, the validity of their findings has not
been confirmed.  Therefore, it is worthwhile investigating the
sticking probability for various materials.

In the present study, we performed MD simulations of water molecules
for a wide range of initial supersaturation ratios and temperatures.
We observed the nucleation rates and derived the formation free energy
by using the size distribution of the clusters.  By comparing these
numerical results with theories, we can test the theoretical models.
We also obtained the sticking probability from the MD simulations
based on the previous method\cite{tanaka2011,diemand2013}. In Section
II, we describe the numerical procedure for our MD simulations, and in
Section III we present our numerical results and compare the numerical
results with the theories. We also determine the sticking probability
and the dependence of the supersaturation ratio.  In Section IV, we
summarize the results of the present study. \\

\section{NUMERICAL PROCEDURE AND  ANALYSIS}\label{sec:MD}

We performed simulations of the nucleation process in systems of 4,000
of water molecules, using the SPC/E rigid water model\cite{berendsen}.
We assumed $NVT$ (constant volume and temperature) ensembles and used
a three-dimensional periodic boundary condition.  Initially, the
molecules are located randomly without being overlapped with others
and the system is relaxed to be in the equilibrium state at 1000 K.
Using this initial condition, we started the simulations.  In our
simulations, the number of water molecules was not so large.  In order
to improve the statistical accuracy, we performed 20 runs with
different initial positions and velocities of molecules for each
parameters of temperature $T$ and initial supersaturation ratio $S_0$.
The time step was set as 2.0 fs.  The total number of time steps was
$(4$-$10) \times 10^6$ in each run.  The simulation box contained 4000
water molecules and 4000 (or 8000) argon carrier gas molecules to
control the temperature.  To compute the long-range electrostatic
interactions, we adopted the reaction field-zero
method\cite{schulz2009} in which a dielectric constant of infinity and
a group-based cutoff are employed.  In this study we set the cutoff
radius to 3.5 nm.

The computational region is a cube with periodic boundaries. By
varying the box size $L$, we set the initial number density of
molecules or the initial supersaturation ratio $S_0$.  The
supersaturation ratio is defined as $S=P_1/P\sub{sat}$, where $P_1$
and $P\sub{sat}$ are the partial pressure of the monomers and the
pressure in a saturated vapor.  The resulting supersaturation is
approximately given by $S=P\sub{total}/P\sub{sat}$ with the total
pressure $P\sub{total}$ because of small amount of clusters larger
than dimers.  Simulations were performed for various temperatures and
supersaturation ratios, {\it i.e.,} $T= 250$-375 K and $S=3$-25.  The
parameter sets for each run are shown in Table I. 28 parameter sets
were chosen, so that the total number of simulations is 560.
For a sufficiently high supersaturation ratio,
the simulation gives a very large nucleation rate
and the number of the vapor molecules decreases so quickly.
In such a case, a steady nucleation with a constant 
supersaturation ratio is not realized.
In our simulations, in order to observe steady nucleation,
we chose lower supersaturation ratios compared with the previous 
study\cite{matsubara2007}.
 For the interaction between carrier 
 gas and water, we used the potential of the Lennard-Jones type: 
\begin{eqnarray}
V(r)= 4 \varepsilon [ (\sigma/r)^{12}-(\sigma/r)^6],
\end{eqnarray} 
where $r$ is the distance between argon and oxygen and
 the parameters  $\sigma$ and $\varepsilon$ are given by 
$\sigma=\sqrt{\sigma\sub{Ar}\sigma\sub{O}}$ 
 and $\varepsilon=\sqrt{\varepsilon\sub{Ar} \varepsilon\sub{O}}$
 with 
$\sigma\sub{Ar}=3.40 \AA$, $\sigma\sub{O}=3.17 \AA$,  
$\varepsilon\sub{Ar}=0.979$ kJ mol$^{-1}$, and 
$\varepsilon\sub{O}=0.652$ kJ mol$^{-1}$~\cite{jorgensen}. 
The carrier gas temperature $T$ was 
controlled directly with a Nos\'e-Hoover thermostat, while the
temperature of the water was controlled indirectly through interaction
with the carrier gas molecules.  The temperature of the water
molecules $T\sub{w}$ deviates from the temperature
of the carrier gas because of the latent heat effect that results from
condensation formation.  We defined clusters using the bonding
criterion that the interaction energy of a molecular pair is less than
-10 kJ/mol\cite{matsubara2007}.

We obtained the nucleation rate and compared the number density of
clusters based on the same method as in the previous
studies\cite{tanaka2011,diemand2013}.  We also derived the formation
free energies from the cluster size distribution\cite{tanaka2014}.  To
derive the formation free energy of a cluster $\Delta G_i$, where $i$
is the number of molecules in the cluster, we use the relation between
the equilibrium size distribution $n\sub{e}(i)$ and the formation free
energy of a cluster $\Delta G_i$\cite{tanaka2011,diemand2013}:
\begin{eqnarray}
{\Delta G_i \over kT} = \ln \left( n(1) \over n\sub{e}(i) \right),
\label{delta-g} 
\end{eqnarray} 
where $n(1)$ is the number density of the monomers.

The steady nucleation rate $J$ is the net number of the transition from 
$i$-mer to $i+1$-mer and given by 
\begin{eqnarray}
J=R^{+}(i)n(i)-R^{-}(i+1)n(i+1),
\label{steadyj} 
\end{eqnarray} 
where $R^{+}(i)$ is the transition rate from a cluster of $i$
molecules, $i$-mer, to ($i$+1)-mer per unit time, {\it i.e.,} the
accretion rate, and $R^{-}(i)$ is the transition rate from $i$-mer to
($i$-1)-mer per unit time, {\it i.e.,} the evaporation rate of
$i$-mer.  $R^{+}(i)$ is given by
\begin{eqnarray}
 R^+(i) =\alpha n(1) v\sub{th}  (4\pi r_0^2 i^{2/3}),
\label{rate} 
\end{eqnarray}
where $v\sub{th}$ is the thermal velocity, $\sqrt{kT/2 \pi m}$, and
$r_0$ is the radius of a monomer, $(3 m / 4 \pi \rho\sub{m})^{1/3}$
where $m$ is the mass of a molecule and $\rho\sub{m}$ is the bulk
density.  The evaporation rate is obtained from the principle of
detailed balance in thermal equilibrium.
\begin{eqnarray}
 R^{-}(i+1)n\sub{e}(i+1)=R^{+}(i)n\sub{e}(i).   
\label{tranm} 
\end{eqnarray}
  From Eqs.(\ref{rate}) and (\ref{tranm}), we have 
\begin{eqnarray}
J=R^{+}(i)n(i)-R^{+}(i)n\sub{e}(i) {  n(i+1) \over n\sub{e}(i+1) },
\label{steadyj2} 
\end{eqnarray}
which leads to
\begin{eqnarray}
{  n\sub{e}(i) \over n(i)  } = { n\sub{e}(i-1)  \over n(i-1) }
\left( 1- {J \over R^{+}(i-1)n(i-1)} \right)^{-1}. 
\label{neq} 
\end{eqnarray}
Equation~(\ref{neq}) shows $n\sub{e}(i)$ agrees well with the number
density of the clusters $n(i)$ obtained by the MD simulations for
subcritical clusters $i \le i^*$, because $J \ll R^{+}(i)n(i)$ for $i
\le i^*$. From Eq.(\ref{neq}), we obtain $n\sub{e}(i)$ if  $J,
n(i)$ and $n\sub{e}(i-1)$ are given.  For example $n\sub{e}(2)$ is obtained by
$J, n(2),$ and $n\sub{e}(1) [=n(1)]$. Also for a larger $i$,
 we can obtain $n\sub{e}(i)$ recursively. 
 Matsubara et
al.\cite{matsubara2007} calculated the equilibrium number density, 
using the different formula derived by Yasuoka and
Matsumoto\cite{ym1} from Fokker-Plank equation.
% in which the cluster size $i$ was treated
%as the continuous variable and the summation from $i=1$ to $i=i$ was
%expressed by the integral. 
 Our formula
 leads to the same  as Matsubara et al.\cite{matsubara2007}
 for $i^{*}>>1$ (see Appendix).
%So the obtained formula was valid for $i >>1$.
% Since the sizes of critical clusters are considerably small, we use
%  (\ref{neq}) which leads the evaluation of $n\sub{e}(i)$ to higher
%  accuracy for smaller values of $i$.
% In their formula the large size of critical cluster 

%we can estimate the values of the
%formation free energy for a subcritical cluster 
%with $i \le i^*$. 
In the theoretical models, the formulae for $\Delta G_i$ are given by
\begin{eqnarray}
{\Delta G_{\rm CNT, i} \over kT} &=& 
-i \ln S + \eta i^{2/3}, \nonumber \\
{\Delta G_{\rm MCNT, i} \over kT} &=& 
-(i-1) \ln S + \eta (i^{2/3}-1), \nonumber \\
{\Delta G_{\rm SP, i} \over kT} &=& 
-(i-1) \ln S + \eta (i^{2/3}-1)+ \xi(i^{1/3}-1), 
\label{delta-g-theor} 
\end{eqnarray}
where $\Delta G_{\rm CNT, i}$, $\Delta G_{\rm MCNT,i}$, and $\Delta
G_{\rm SP,i}$ are the formation free energies of a cluster in the CNT,
the modified CNT (MCNT), and the SP model, respectively, and $\eta$
and $\xi$ are parameters determined by the bulk surface energy and
second virial coefficient\cite{tanaka2011,diemand2013}. 
 In the classical theory, there is a theoretical inconsistency in the $S$
dependence.  Consequently the MCNT has often been used.  In the next section, 
 we compare the CNT, MCNT and SP models with the our numerical results. 
 Although there are some differences between 
 the thermodynamic quantities
 obtained by using SPC/E force  and those by experiments of real water,  
 we should use those of SPC/E force for the
 consistency to the model. 
 For thermodynamic quantities such
as the surface tension and the saturated vapor pressure of the SPC/E
water, we use the same data as Matsubara et al.\cite{matsubara2007}
  \\

\section{RESULTS}

\subsection{A typical case}

\begin{figure}
\includegraphics[height=.3\textheight]{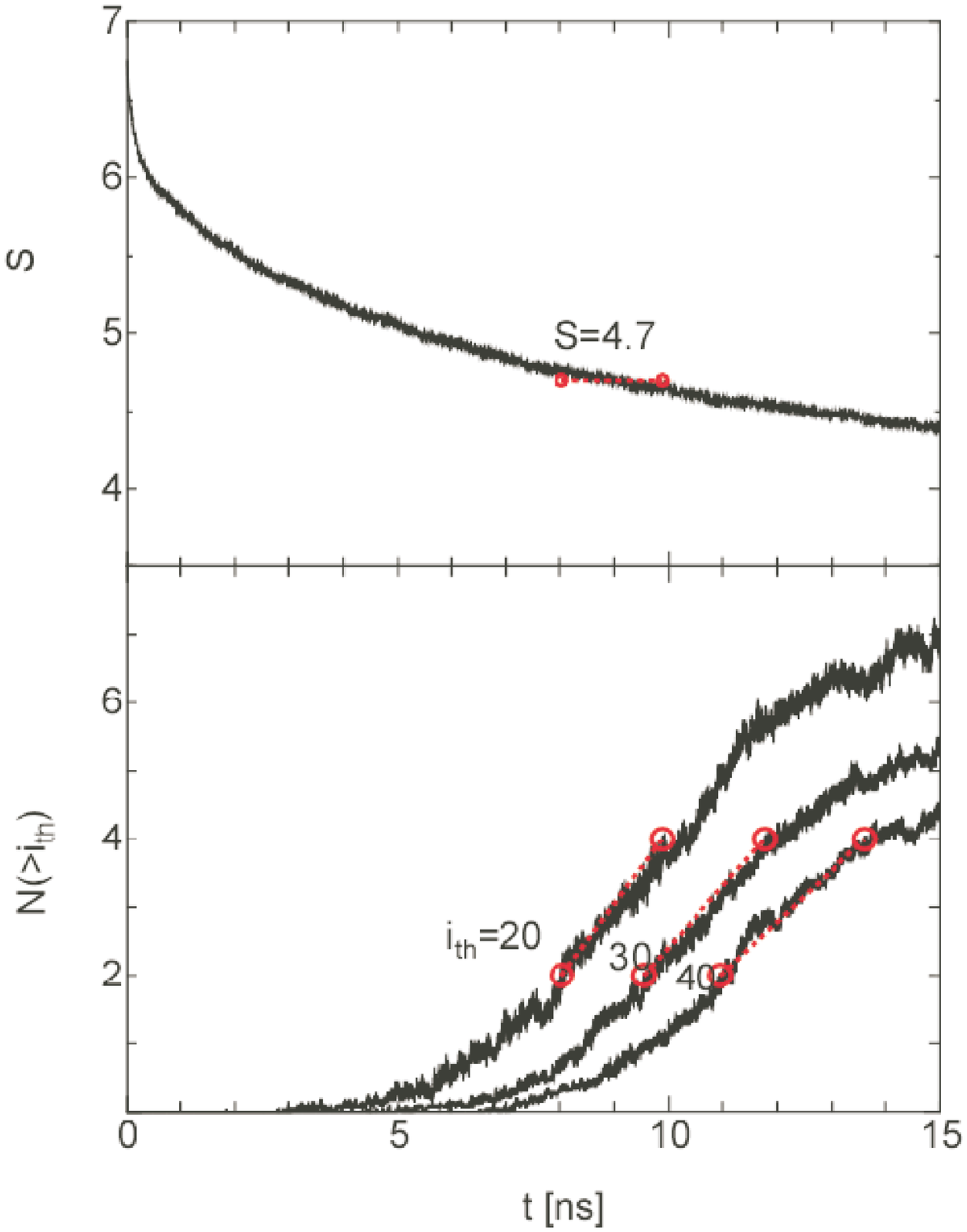}
\caption{ Supersaturation ratio and  
number of clusters above various threshold sizes as 
a function of time for run 5d, where $N(>i\sub{th})$ is
 averaged from 20 runs with the same initial conditions. 
}\label{jslope-0.3}
\end{figure}

Let $N(> i\sub{th})$ denote the number of clusters larger than a
threshold size $i\sub{th}$.  Based on Yasuoka and Matsumoto
method\cite{ym1}, we can obtain the nucleation rate from $N(>
i\sub{th})$.  Figure~\ref{jslope-0.3} shows time evolution of the
supersaturation ratio and the number of clusters, $N(> i\sub{th})$, in
a typical case where $T = 350$ K and $S_0 = 6.67$ (run 5d in Table I).
Note that $N(> i\sub{th})$ is averaged over 20 runs with different
initial positions and velocities of molecules.  In
Figure~\ref{jslope-0.3}, we plot $N(> i\sub{th})$ for $i\sub{th} =$20,
30, and 40.

The threshold $i\sub{th}$ should be larger than the
critical size $i^*$ for the evaluation of the nucleation rate.
 In this case, the critical size $i^*$ is
estimated to be 14 (or 7) with the SP (or CNT) model.  In this figure,
the number of stable nuclei $N(> i\sub{th})$ increases almost linearly
in the period of $t= (6-12)$ns.  
From the slope, the nucleation
rate is measured with $J(=\dot{N}[> i\sub{th}]/L^3)$.
In the evaluation of J, we chose the slope between
the times at $N(> i\sub{th})=2$ and 4, {\it e.g.,}
 $t =8-10$ ns for $i\sub{th}$=20.
Then, the nucleation rate is obtained as
 $8.92 \times 10^{24}$,  $7.37 \times 10^{24}$, 
 and  $6.19 \times 10^{24}$cm$^{-3}$s$^{-1}$, 
for $i\sub{th} =20$, 30, and 40, respectively.
In this evaluation of the nucleation rate, uncertainties of 20-30\% 
exist, depending on the choice of $i\sub{th}$ and the time interval.

 During this period of $t= (8$-$10)$ ns, the average value of the
supersaturation $S$ and the temperature of the water molecules
$T\sub{w}$ were $4.7$ and $361$~K, respectively.  Using these values,
we can calculate $J$ predicted by the theoretical models.  In the
MCNT, the CNT and the SP models, the nucleation rate are obtained as
$1.5 \times 10^{27}$, $8.0 \times 10^{25}$, and $J=2.1 \times
10^{24}$cm$^{-3}$s$^{-1}$, respectively.  The prediction of the SP
model is much closer to the values obtained in the MD simulations than
the MCNT and the CNT.  \\

\subsection{Nucleation rate}

Fig.~2 shows the time evolutions of $N(>i\sub{th})$ for $i\sub{th}=$
20, 30, and 40 in the various runs (6e, 5c, 3c and 2e).  We evaluate
the nucleation rates from the slopes between the times at $N(>
i\sub{th})=2$ and $N(> i\sub{th}) =4$ for all runs except run 6e. For
run 6e, we use the slope between the times at $N(> i\sub{th})=1$ and 3
 because of the small number of clusters.  The
nucleation rates obtained in all MD simulations are listed in Table
II, where we chose $i\sub{th}=$20 for all runs because the size of
critical cluster is estimated to be less than 20 for all runs (see
Section III.D).  As stated in the previous section, there are some
uncertainties in the estimation of the nucleation rates. So we include
the errors in the nucleation rates in Table II.

Fig.~\ref{J-S} shows the nucleation rates obtained by the MD
simulations as a function of the supresaturation ratio for various
temperatures.  The previous results are also shown: crosses and
triangles are the results by Matsubara et al.\cite{matsubara2007} and
Zipoli et al.\cite{zipoli}, respectively.  The nucleation rates by SP
model (solid curves) are also plotted.  The obtained nucleation rates
are about one-order-of-magnitude smaller than Matsubara et
al.\cite{matsubara2007} because of the smaller supersaturation ratios.
We find good agreements between the MD simulations and the SP model
within one-order-of-magnitude in the cases of low temperatures for
both our runs and the previous results.

Figure~\ref{ratio-j} shows the ratios of the nucleation rates in the
MD simulations (at $T\sub{w}$) to the rates obtained with the
theoretical models. The results are plotted with circles for the SP
model, triangles for the MCNT, and crosses for the CNT.  The
nucleation rates predicted by the SP model are within one order of
magnitude of the MD simulations, whereas the nucleation rates
predicted by the CNT and MCNT are much larger than the MD simulations.
We also plotted the ratios $J\sub{MD} / J\sub{SP}$, by using results
by Matsubara et al.\cite{matsubara2007} (i.e., data in Table I in
\cite{matsubara2007}).  Our results agree well with theirs.  The ratio
of $J\sub{MD} / J\sub{SP}$ gradually increases with temperature in
both results.  The nucleation rates obtained with the theoretical
models are also listed in Table II.

In Figure~\ref{ratio-j}, we also show the
results of the Lennard-Jones system\cite{tanaka2011} for reference.  
The SP model also predicts the nucleation rates to
within one order of magnitude for the Lennard-Jones system, although
 the supersaturation ratios are limited to be small.  
 In the case of the Lennard-Jones
system, the CNT considerably underestimates the nucleation rates (
{\it i.e.,} the $J\sub{MD} / J\sub{CNT}$ ratio is much larger than $10^{9}$)
at $T<0.4 (\varepsilon/k\sub{B})$, where $\varepsilon$ and $k\sub{B}$
are the depth of the Lennard-Jones potential and the Boltzmann
constant, respectively.  On the other hand, the CNT predictions agree
with the nucleation rates obtained by the MD simulations to within two
orders of magnitude in the case of water.  

In Fig.~\ref{jscale}, we show the nucleation rates as a function of
 $\ln S /(T_c/T-1)^{1.5}$ in order to compare our results with the
 scaling relation proposed by Hale\cite{hale}.  The data by the MD
 simulations and the theoretical predictions are the same as those in
 Fig.~\ref{J-S}.  For the MD simulations using SPC/E water model, we
 set $T_c=630$ K for the consistency to the model\cite{matsubara2007}.
 We also put the experimental data for water nucleation
 \cite{miller,wolk,khan,kim, brus2009, manka}.
  For the experimental results and 
 Zipoli et al.\cite{zipoli}, $T_c$ is set to be 647 K\cite{wolk}. From
 Fig.~\ref{jscale}, $J$ seems to be scaled by $\ln S
 /(T_c/T-1)^{1.5}$, but is not proportional to it.  \\

\subsection{Cluster size distributions}

Figure~\ref{ni-total-1} shows the size distributions $n(i)$ obtained
from 8 runs at various temperatures $T\sub{w}$ and initial
supersaturation ratios.  For each temperature, we show high-$S$ cases
(right column) and low-$S$ cases (left column).  The MCNT
significantly overestimates the distributions (by factors $10^{3}$).
In all cases, the SP model predicts the size distributions more
accurately than both the MCNT and the CNT for clusters smaller than
the critical size.  At $i=2$ the SP model agrees perfectly with the MD
simulations in all cases.  However, the SP model disagrees with the
results of the MD simulations for large clusters.  There is also a
relatively large deviation at $i=4$.  This is due to the stable
four-ring structure\cite{merikanto2007, qian}.  These results for the size
distributions are consistent with the nucleation rate, as shown
Section II.B. \\

\subsection{Formation free energy}

We derive the formation free energy of a cluster $\Delta G_i (S)$,
using the equilibrium size distribution $n\sub{e}(i)$ given by
Eq.(\ref{neq}). In the derivation, we use the nucleation rate $J$ and
$n(i)$ obtained by the MD simulations and $R^{+}(i)$, where we use the
value of $\alpha$ measured by the simulations (see Section II.E. and
Table II.)  Using the $S$-dependence of $\Delta G_i$ in
Eq.(\ref{delta-g-theor}) except the CNT, we also obtain $\Delta G_i$
at $S=1$ as
\begin{eqnarray}
\Delta G_i (S\!=\!1) =  \Delta G_i (S) 
+(i-1) kT \ln S, 
\label{delta-g2} 
\end{eqnarray}
where $\Delta G_i$ is obtained from Eq.(\ref{delta-g}).  $\Delta
 G_i(S\!=\!1)$ corresponds to the contribution of the surface effect
 to the formation free energy of a cluster.  If we know $\Delta
 G_i(S\!=\!1)$, we can predict $\Delta G_i$ and the nucleation rate
 for any value of $S$. 

  Fig.~\ref{eqdist-deltag} shows the size distribution of clusters
 obtained by MD simulations and the equilibrium one in run 5d.  The
 formation free energies $\Delta G_i $ are shown for $S=4.7$ and $S=1$
 in Fig.~\ref{eqdist-deltag}. From $\Delta G_i$ for $S=4.7$, we obtain
 the maximum value of the formation free energy and the size of the
 critical cluster in run 5d, {\it i.e.,} $\Delta G_{i}^{*}=7.0 kT$ and
 $i\sub{*}=12$.  Table II shows the sizes of critical cluster obtained
 by the simulations for all runs.

In Fig.~\ref{deltag-deltagsp}, we compare the maximum value of the
  formation free energy at the critical cluster between the
  theoretical models and the MD simulations for all runs.  Evaluating
  $\Delta G_{i}^{*}$ in the theoretical model, we use the values of
  the temperature $T\sub{w}$ and the supersaturation ratio obtained by
  the simulations shown in Table II.  The CNT and MCNT considerably
  underestimate the maximum free energies.  The SP model predicts the
  maximum free energies to a higher accuracy than the CNT and MCNT:
  the SP model predicts the maximum free energies within 30 \%.  In
  Matsubara et al.\cite{matsubara2007}, the similar comparison was
  performed. There are some differences between the our results and
  \cite{matsubara2007}.  We confirmed that there is an inconsistency
  in Matsubara et al.\cite{matsubara2007} where they used slightly
  smaller temperatures in their figures than the true values of 
  $T\sub{w}$ in Table I.\cite{matsubara2007} 
  (private communications).

Figure~\ref{deltag5} shows the formation free energy $\Delta G_i
(S\!=\!1)$ obtained by simulations at various temperatures.  
Predictions of the theoretical models are plotted with solid
lines (SP model) and dashed lines (MCNT). The SP model exhibits much
better agreement with the numerical results than the MCNT.  The SP
model reproduces the formation free energy of a cluster to within 10 \%.
The CNT also reproduces the formation free energy of a cluster better than
the MCNT, but the gradient of the CNT data is different to the numerical
results.
 
{From $\Delta G_i$ obtained by MD simulations, 
we can calculate the nucleation rate by using 
\begin{eqnarray}
 J= \left\{ 
    \sum_{i=1}^{\infty}
  { 1 \over R^{+}(i) n\sub{e}(i)} \right\}^{-1},
\label{j1}
\end{eqnarray}
where $n\sub{e}(i)$ is given by Eq.(\ref{delta-g}). 
We confirmed the nucleation rates given by Eq.(\ref{j1})
predict $J\sub{MD}$ measured by the MD simulations within the accuracy 
 of 1 \% for all runs.  }\\

\subsection{Sticking probability}

As shown in Section II, the nucleation rate is obtained from the
equilibrium number density $n_e(i)$ and the accretion rate which is
proportional to the sticking probability $\alpha$.  Using the growth
rate of stable clusters obtained in the present MD simulation, we
evaluated the sticking probability.

Tanaka et al.\cite{tanaka2011} evaluated the sticking probability from the
growth rate of stable clusters, $di/dt$, neglecting the effect of
evaporation.  In the present study the evaporation effect was taken
into account based on the method of Diemand et al.\cite{diemand2013}.
  The growth
rate of clusters, $di/dt$, is given by
\begin{eqnarray}
\frac{di}{dt} = R^+(i) -R^-(i).
\label{growthrate}
\end{eqnarray}
  From Eqs.(\ref{rate}) and (\ref{tranm}), we have
\begin{eqnarray}
\frac{di}{dt} = 4\pi r_0^2 i^{2/3} \alpha n(1) v\sub{th} 
                \left[ 1- \left(i-1 \over i \right)^{2/3} 
 e^{\Delta G_{i}-\Delta G_{i-1} \over kT} 
   \right],
\end{eqnarray}
where we use $n\sub{e}(i)=n(1) \exp[-\Delta G_i / (kT)]$.

As shown in Eq.(\ref{delta-g-theor}), $\Delta G_i$ has a bulk term
($\propto i$) 
 and a surface term $(\propto i^{2/3})$. Since the other terms in
Eq.(\ref{delta-g-theor}) are smaller than these  terms,
 the sum, $\Delta
G_i + i kT \ln S$ is approximately
 proportional to $i^{2/3}$  and 
\begin{eqnarray}
\Delta G_{i}-\Delta G_{i-1} \simeq - kT \ln S   
\end{eqnarray}
for $i \gg 1$. Hence 
for $i \gg 1$, we obtain   
\begin{eqnarray}
\frac{di}{dt}
 =\alpha n(1) v\sub{th}  (4\pi r_0^2 i^{2/3})
                \left[ 1- {1 \over S } 
   \right].
\label{didt}
\end{eqnarray}
%\begin{eqnarray}
%\frac{di^{1/3}}{dt}
% =\alpha n(1) v\sub{th}  ({4\pi r_0^2  \over 3})
%                \left[ 1- {1 \over S } 
%   \right].
%\label{didt}
%\end{eqnarray}
Equation~(\ref{didt}) suggests that the evaporation becomes
considerable when the supersaturation ratio is small.  Accordingly,
the sticking probability is given by
\begin{eqnarray}
\alpha  = { 3 \over 4\pi r_0^2 v\sub{th} n(1) }\left(1- {1 \over S}
 \right)^{-1}
\frac{di^{1/3} }{dt}. 
\label{alpha_s}
\end{eqnarray}
The time derivative of $i^{1/3}$ was evaluated from the MD
simulations. Figure~\ref{rad-t} shows the time derivative of $i^{1/3}$
for run 5d.  Based on the slope, the sticking probability was measured
to be $0.472$.  Using Eq.(\ref{alpha_s}), we obtained the
sticking probability for all runs.  The values of $\alpha$ are also
listed in Table II.  The results are shown in
Figure~\ref{alpha-fit}~(a) as a function of supersaturation rate.  By
plotting a line of best fit, we obtained an empirical formula for the
sticking probability:
\begin{eqnarray}
\ln \alpha = 1.16 \ln S \left( T \over 273 \mbox{K} \right)^{3.3}-5.3
\label{sticking}
\end{eqnarray}
 (see Fig.~\ref{alpha-fit}~(b)).  The sticking probability increases
with the supersaturation ratio, which is consistent with the
Lennard-Jones system\cite{tanaka2011,diemand2013}.   The formula
(\ref{sticking}) is valid within the parameter range of $S$ in which
our MD simulations are done.  In this range ($S > 3$), 
Eq.(\ref{sticking}) indicates that $\alpha$ increases
with the temperature for a fixed $S$. 
 In the case of the flat surface and
the equilibrium state ($S=1$), on the other hand, the decreasing of
$\alpha$ with the increasing temperature is reported by the previous
study\cite{matsumoto1996}.  Around the equilibrium state
 ($S \simeq 1$),  the sticking probability 
 should be examined  in the future work.

We evaluated $\alpha$, by using the growth rate of clusters 
larger than the critical size with Eq.(\ref{alpha_s}). 
Matsubara et al.\cite{matsubara2007} 
 evaluated the sticking probability (or the
sticking coefficient), by directly measuring the condensation flux  
 ({\it i.e.,} accretion rate)
 for clusters smaller than or comparable to the critical size
instead of the growth rate.  Their values are one-order-of-magnitude
greater than ours.  This would indicate that the sticking probability
might be different between the under-critical and the over-critical
sizes.  Matsubara et al.\cite{matsubara2007} 
 suggested that their large sticking
coefficients are due to the large surface area of clusters.  For small
clusters, larger surface areas than the simple prediction with the
bulk density are also reported by other studies\cite{ym1,raymond}.
A high supersaturation ratio in Matsubara et al.\cite{matsubara2007} 
can be another factor which causes the above difference in $\alpha$. 

% By directly measuring the condensation flux $k^+$ for clusters
% with relatively small size ($\siml i_*$), 
% Matsubara et al.\cite{matsubara2007} 
% evaluated the sticking coefficient with 
% $k^+/ [ n(1) v\sub{th} A_i] $, where $A_i$ is the surface area of
% sphere with the bulk density.  Their sticking coefficients
% are one-order-of-magnitude greater than our
% evaluated values.  They suggested that the large sticking coefficient
% comes from the large surface area of clusters. The larger surface area
% than that of sphere with the bulk density 
%is also reported  for small clusters  by other studies 
% The difference of $\alpha$ between our study and \cite{matsubara2007} may
%come from the different conditions such as the
%cluster size and supersaturation ratios.   } 

% From our method, we can obtain the sticking probability 
% as a function of the averaged monomer number density and
% thermal velocity. On the other hand, 
% there is an another way to
% to measure  {\bf the incident flux  and 
% $R^+(i)$ directly from the  MD simulations.
% If we use the apparent values by the  MD simulations,} 
% we get more realistic sticking probabilities which 
% may change from those  obtained by our method. 
% The relation between two quantities 
% should be examined in a future work. 

\section{SUMMARY AND FUTURE WORK}\label{sec:summary}

Our results on the nucleation of water molecules
are summarized as follows.

1.  The cluster size distributions, the formation free energy of a
cluster, and the nucleation rates obtained from the MD simulations are
compared with the predictions by the theoretical models (see
Figures~3-6 and Table II). The CNT and the MCNT
overestimate the nucleation rates (or the number density of critical
clusters) for all runs. 
On the other hand, the SP model gives
a better prediction. The nucleation rates predicted by the SP model 
are within one order of magnitude. 
The free energy of cluster formation $\Delta G_i$
(or the cluster integrals) for small clusters ($i \siml 20$)
 is also well predicted by the SP model 
(with 10\% accuracy) for a wide range of temperature.
In this study, only large supersaturation ratios and nucleation rates
were considered.  In order to verify the validity of the SP model for
smaller supersaturation ratios, simulations using a larger number of
molecules are necessary.

2.  The sticking probability of vapor molecules onto
clusters was measured in supersaturated conditions, 
by using the growth rate of 
stable clusters in the MD simulations.
 The sticking probability $\alpha$ increases with the
supersaturation ratio.  This tendency is the same  as exhibited by the
Lennard-Jones system\cite{tanaka2011,diemand2013}.  
The lowest value of $\alpha$ was observed to be 0.3 for
$T \simeq 380$ K and $S \simeq 3.3$. 
It is expected that $\alpha$ would be much smaller than 
0.3 for lower supersaturation ratios.
For such low-$S$ cases, we must estimate $\alpha$ precisely for the
evaluation of $J$.  
\\

\section{Acknowledgments}

 We are grateful to Dr. Matsubara for discussions and valuable
 comments.  H. Tanaka and K. K. Tanaka acknowledge Dr. Diemand and
 Dr. Angelil for fruitful discussions.  We would like to thank the
 anonymous reviewers for their valuable suggestions to improve the
 quality of the paper.  This work was supported by the Japan Society
 for the Promotion of Science (JSPS).  \\

\section{Appendix}

 In the present paper, we use 
Eq.(\ref{neq}) to obtain the equilibrium number density. 
In this appendix, we show the relation Eq.(\ref{neq}) to 
the formula used in 
Matsubara et al.\cite{matsubara2007}.
Equation~(\ref{neq}) is a  recurrence relation and yields
\begin{eqnarray}
{  n\sub{e}(i) \over n(i)  } &=& \prod_{k=1}^{i-1}
\left( 1- {J \over R^{+}(k)n(k)} \right)^{-1} \nonumber \\
 &=& \prod_{k=1}^{i-1}
\exp \left\{ -\ln \left( 1- {J \over R^{+}(k)n(k)} \right) \right\}.
\label{neq2} 
\end{eqnarray}
If we assume 
\begin{eqnarray}
 {J \over R^{+}(i)n(i)}   \ll 1,
\label{neq3} 
\end{eqnarray}
for all $i$,
we obtain 
\begin{eqnarray}
{  n\sub{e}(i) \over n(i)  } 
& = & \prod_{k=1}^{i-1}
\exp \left\{ -{J \over R^{+}(k)n(k)} \right\} \nonumber \\ &=&
\exp \left\{ -J \sum_{k=1}^{k-1}{1 \over R^{+}(k)n(k)} \right\}
\label{neq4} 
\end{eqnarray}
which  corresponds to the formula (27) in Matsubara et al\cite{matsubara2007}.
Since the nucleation rate is expressed with the Zeldovich factor $Z$:
\begin{eqnarray}
J
& \simeq & R^+(i_*) n\sub{e}(i_*) Z, \nonumber \\
Z& =& 
 \left\{ { -1 \over 2 \pi k T}  \left(
 d^2 \Delta G_i \over d i^2 \right)_{i=i_*} 
  \right\}^{1/2}, 
\label{neq5} 
\end{eqnarray}
the condition (\ref{neq3}) is written as
\begin{eqnarray}
 Z  \ll  1.
\label{neq6} 
\end{eqnarray}
In the theory of CNT, MCNT, and SP model,
the Zeldovich factor is given by 
\begin{eqnarray}
 Z  = {1 \over 3 } i_*^{-2/3} \sqrt{ {\eta \over \pi }
 + {\xi \over \pi}i_*^{-1/3} }, 
\label{neq7} 
\end{eqnarray}
where $Z$ in the CNT (or MCNT) corresponds to the case of $\xi=0$. 
Equation~(\ref{neq7}) indicates that
 the condition (\ref{neq3}) is satisfied in the case of  
 $ i^{*} \gg (\eta / 9 \pi)^{3/4} \sim 1$. 
\\
 


\bibliographystyle{aipproc}   % if natbib is available

%%%%%%%%%%%%%%%%%%%%%%%%%%%%%%%%%%%%%%%%%%%
%% The following lines show an example how to produce a bibliography
%% without the help of the BibTeX program. This could be used instead
%% of the above.
%%%%%%%%%%%%%%%%%%%%%%%%%%%%%%%%%%%%%%%%%%%

\newpage

\begin{figure}
\includegraphics[height=.5\textheight]{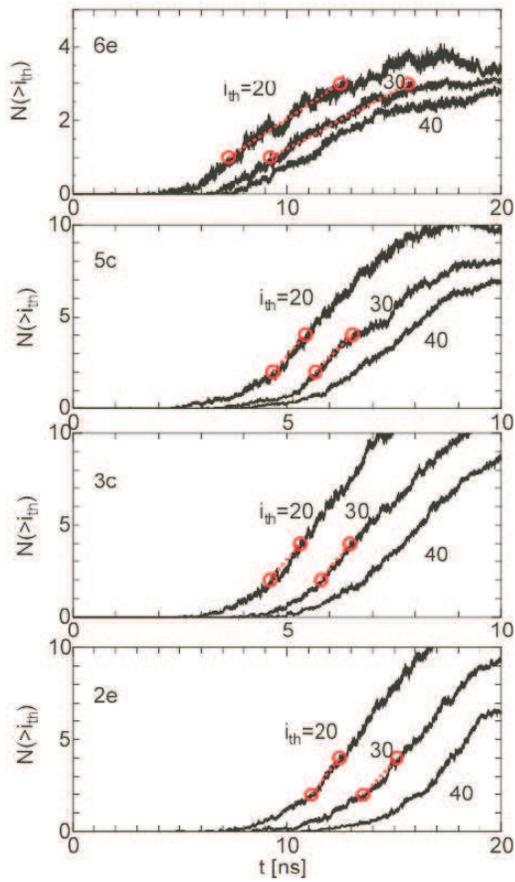}
\caption{ 
Number of clusters above various threshold sizes as 
a function of time for various runs, where $N(>i\sub{th})$ is
 averaged from 20 runs with the same initial conditions. 
}\label{jslope-all}
\end{figure}

\begin{figure}
\includegraphics[height=.25\textheight]{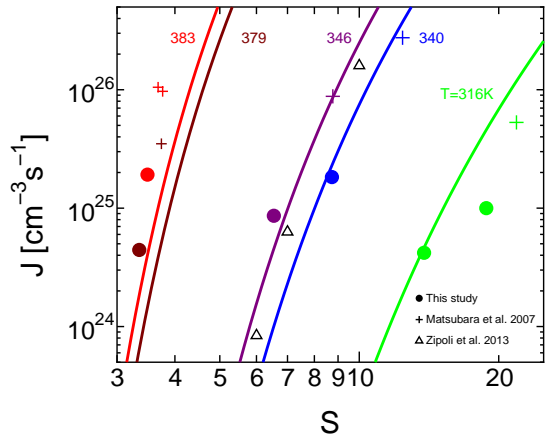}
\caption{ The nucleation rates 
 obtained by the MD simulations as a function of the supresaturation ratio
 for various temperatures. 
 The previous results are also shown: crosses and triangles 
 are the results
 by Matsubara et al.\cite{matsubara2007} 
and Zipoli et al.\cite{zipoli}, respectively.
 In   Zipoli et al.\cite{zipoli},
 we use the nucleation rates at 350~K. 
 The nucleation rates by SP model (solid curve)
 are also plotted.}
\label{J-S}
\end{figure}

\begin{figure}
\includegraphics[height=.5\textheight]{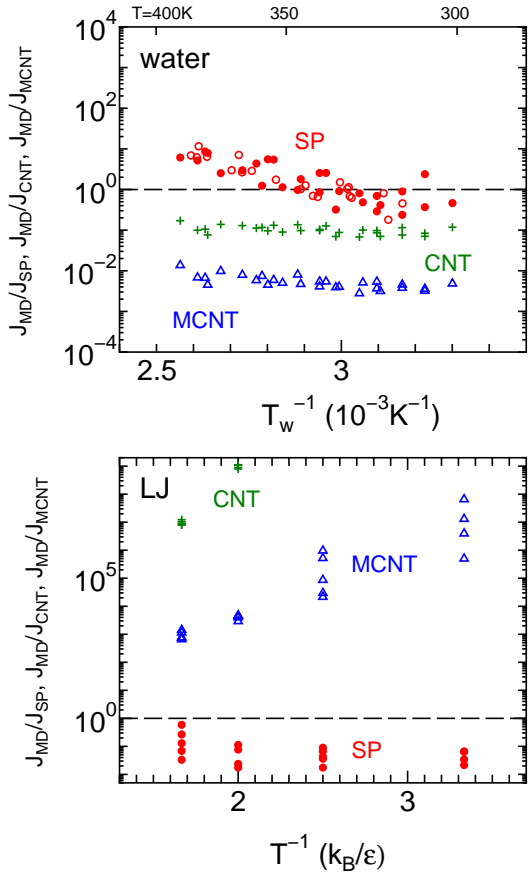}
\caption{  The top panel shows the results
for water obtained in this study.
 The ratios of nucleation rates of our MD simulations
and the theoretical models, plotted with  closed circles for the SP
model ($J\sub{MD} / J\sub{SP}$), triangles for the MCNT ($J\sub{MD} /
J\sub{MCNT}$), and crosses for the CNT ($J\sub{MD} / J\sub{CNT}$).
 The temperature is $T\sub{w}$. 
 By using $J\sub{MD}$ obtained by 
 Matsubara et al\cite{matsubara2007}, 
 we also plot $J\sub{MD} / J\sub{SP}$ with open circles.
  The results of the 
 Lennard-Jones system\cite{tanaka2011}
 are also shown in the bottom panel for reference.
}
\label{ratio-j}
\end{figure}

\begin{figure}
\includegraphics[height=.25\textheight]{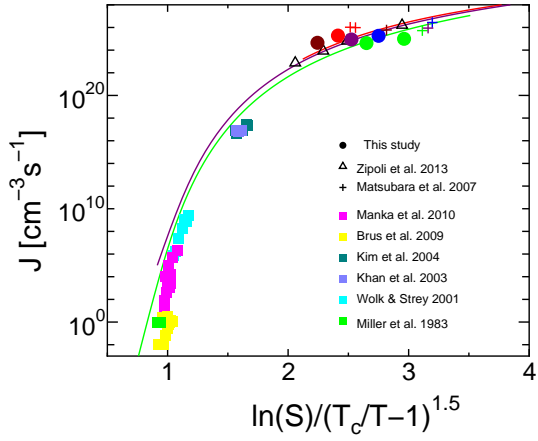}
\caption{ The nucleation rates as a function of 
 $\ln S /(T_c/T-1)^{1.5}$.
  The data by the MD simulations and  
 the theoretical predictions by the SP model (316~K, 
 346~K, and 383~K) are the same as those in Fig.~\ref{J-S}.
 For the MD simulations using SPC/E water model, 
 we set  $T_c=630$ K\cite{matsubara2007}. 
 We also put the experimental data for nucleation of water
 by Miller et al.\cite{miller}, W\"olk and  Strey\cite{wolk},
 Khan et al.\cite{khan}, Kim et al.\cite{kim},
Brus et al.\cite{brus2009}, and Manka et al.\cite{manka}. 
 For the experimental results and  Zipoli et al.\cite{zipoli},
 $T_c$ is set to be 647 K\cite{wolk}. 
  }
\label{jscale}
\end{figure}

\begin{figure}
\includegraphics[height=.45\textheight]{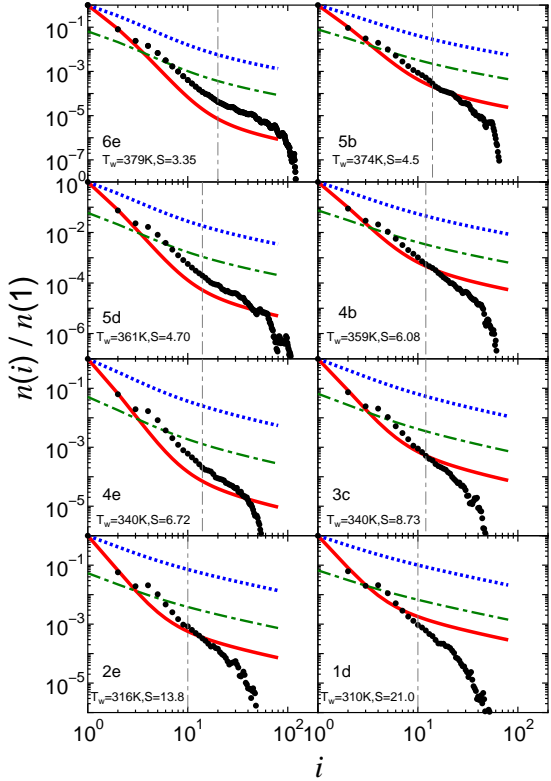}
\caption{ The size distribution of clusters $n(i)$ obtained by the MD
 simulations is indicated by closed circles for the typical case. 
 The size distributions obtained by 
 the SP model (solid curve), the MCNT (dotted curve),
 and the CNT (dotted-dashed curve) are also plotted.
 The dotted-dashed lines show the size of critical clusters 
 evaluated by the MD simulations. }
\label{ni-total-1}
\end{figure}

\begin{figure}
\includegraphics[height=.32\textheight]{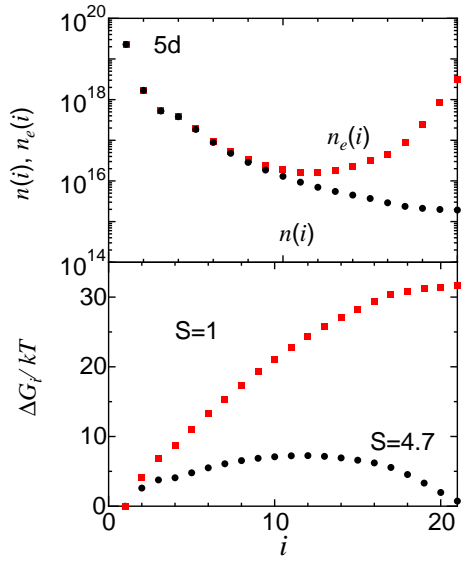}
\caption{ The size distribution of clusters obtained by MD simulations
 (filled circles) and the equilibrium one (filled squares) in run 5d
 are shown in the top panel.  In the bottom panel, the formation free
 energies of a cluster $\Delta G_i (S\!=\!1)/kT$ are shown for 
 $S=4.7$  (filled circles) and $S=1$ (filled squares). }
\label{eqdist-deltag}
\end{figure}

\begin{figure}
\includegraphics[height=.30\textheight]{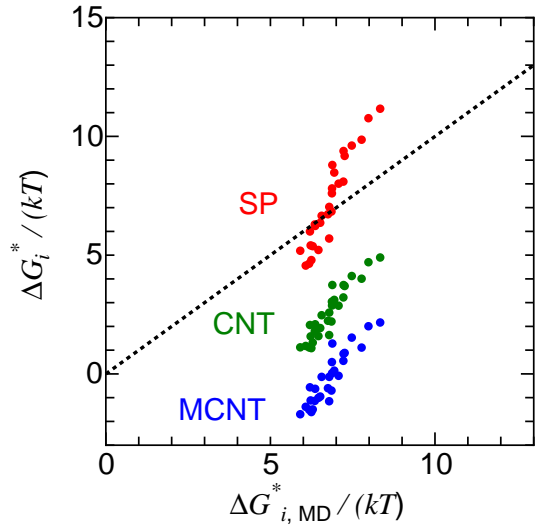}
\caption{ The  comparison of the maximum values of the formation free energy
 obtained by theoretical models 
%($\Delta G\sub{SP,i}^{*}, \Delta G\sub{MCNT,i}^{*}$, and $\Delta G\sub{CNT,i}^{*}$)
 and MD simulations $\Delta G^{*}\sub{i,MD}$. 
}
\label{deltag-deltagsp}
\end{figure}

\begin{figure}
\includegraphics[height=.28\textheight]{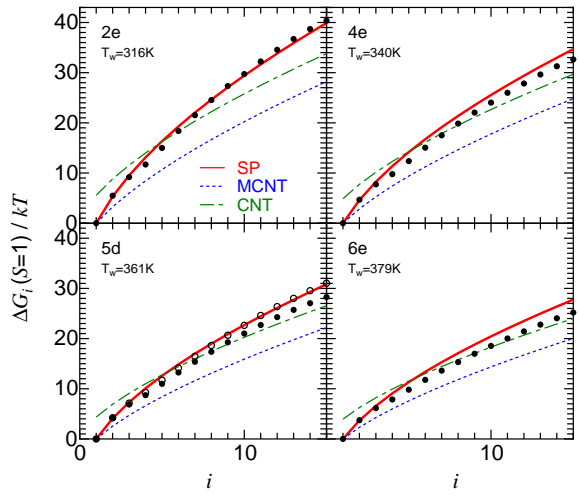}
\caption{ The formation free energy of a cluster $\Delta G_i (S\!=\!1)/kT$
 at various temperatures is shown by the filled circles.
 $T\sub{w}=316$~K (run 2e), 340~K (4e), 361~K (5d), and 379~K (6e).  The
 values obtained by the SP model (solid lines), the MCNT (dotted
 lines), and the CNT (dotted-dashed lines) are also shown. In run 5d, we also plotted the results of $T\sub{w}=359$K (run 4b) with
 open circles.  }
\label{deltag5}
\end{figure}

\begin{figure}
\includegraphics[height=.32\textheight]{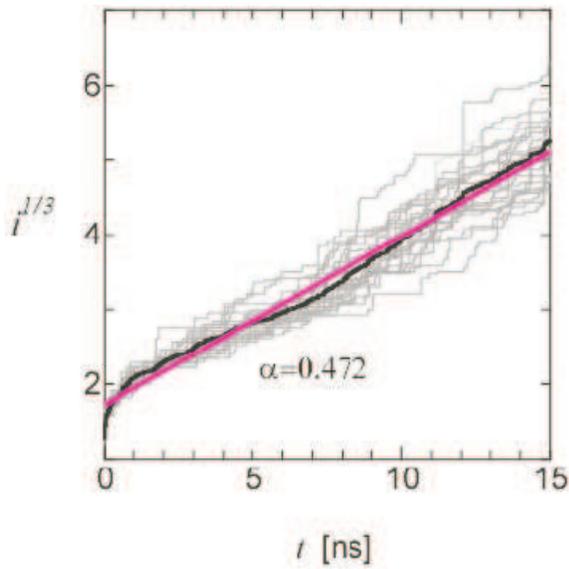}
\caption{ Time evolution of the size of the largest cluster for 20
 runs of 5d.  The average value is plotted by a black
line.  Based on the averaged slope (red  line), the sticking
probability was found to be $0.472$.  }
\label{rad-t}
\end{figure}

\begin{figure}%
\includegraphics[height=.36\textheight]{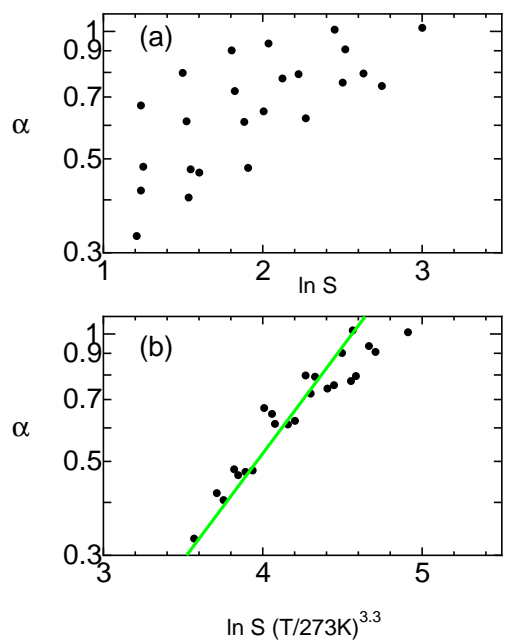}
\caption{ (a) The sticking probabilities obtained by 
 MD simulations as a function of supersaturation. (b)  
 The sticking probabilities as a function of $\ln S (T/273$K$)^{3.3}$
 The equation of the line of best fit is 
$\ln \alpha = 1.16 \ln S (T/273$K)$^{3.3}-5.3$, which is shown by
 a solid line.}
\label{alpha-fit}
\end{figure}

\newpage

{
\begin{table}
\caption{ Parameters in the MD simulations, temperature of
carrier gas $T$, cell size $L$, initial number density of monomers
$n(1)_{t=0}$, number of carrier gas $N\sub{cg}$, and simulation time
$t\sub{end}$.}
\begin{center}
\baselineskip 0.2cm
\renewcommand\arraystretch{0.8}
\begin{tabular}{c  c c c c c } \hline    \hline
   run$\#$    & $T$ [K] & $L$[nm] &  $n(1)_{t=0} [10^{-2}$nm$^{-3}]$ & $N\sub{cg}$ & $t\sub{end}$
 [ns]   \\ \hline  
6e &          375 & 44.10 &4.66 & 4000 & 20 \\
6d &          375 & 42.30 & 5.28 & 4000 & 15 \\
6c &          375 & 40.50 & 6.02 & 4000 & 10 \\
6b &          375 & 36.82 & 8.01 & 4000 & 8 \\
\\
5e &          350 & 52.50 & 2.76& 4000 & 30 \\
5e2 &          350 & 52.50 & 2.76 & 8000 & 30 \\
5d &          350 & 49.50 & 3.30 & 4000 & 15 \\
5c &          350 & 45.00 & 4.39 & 4000 & 10 \\
5b &          350 & 40.91 & 5.84 & 4000 & 8 \\
\\
4e &          325 & 59.90 & 1.86 & 4000 & 20 \\
4e2 &          325 & 59.90 & 1.86 & 8000 & 20 \\
4d &          325 & 54.45 & 2.48 & 4000 & 15 \\
4c &          325 & 49.50 & 3.30 & 4000 & 10 \\
4b &          325 & 45.00 & 4.39 & 4000 & 8 \\
\\
3e &          300 & 65.34 & 1.43& 4000 & 20 \\
3e2 &          300 & 65.34 & 1.43 & 8000 & 20 \\
3d &          300 & 59.40 &1.91 & 4000 & 15 \\
3c &          300 & 54.00 & 2.54 & 4000 & 10 \\
3b &          300 & 49.09 &3.38  & 4000 & 8 \\
\\
2e &          275 & 70.79 &1.13 & 4000 & 20 \\
2e2 &          275 & 70.79 & 1.13 & 8000 & 20 \\
2d &          275 & 64.35 & 1.50 & 4000 & 15 \\
2c &          275 & 58.50 & 2.00 & 4000 & 10 \\
v2b &          275 & 53.18 & 2.66  & 4000 & 8 \\
\\
1e &          250 & 76.23 & 0.903  & 4000 & 20 \\
1d &          250 & 69.30 & 1.20  & 4000 & 15 \\
1c &          250 & 63.00 & 1.60 & 4000 & 10 \\
1b &          250 & 57.27 & 2.13  & 4000 & 8 \\
\hline   \hline 
\end{tabular}
\vspace{0.8cm}
\end{center}
\end{table}
}

{
\begin{table*}
\caption{ Summary of the results of the MD simulations.  $T$
[K]: temperature of carrier gas, $T\sub{w}$ [K]: temperature of water
molecules,  $S$: averaged supersaturation ratio in the
nucleation stage, $J\sub{MD} [10^{24}$ cm$^{-3} $s$^{-1}]$: nucleation
rates obtained by the MD simulations, $J\sub{SP}[10^{24}$cm$^{-3}
$s$^{-1}]$: nucleation rates by the SP model,
 $J\sub{MCNT}[10^{24}$cm$^{-3} $s$^{-1}]$: nucleation rates by the MCNT ,
 $J\sub{CNT}[10^{24}$cm$^{-3} $s$^{-1}]$: nucleation rates by the CNT ,
$i^*$: critical cluster size evaluated by the MD simulations, 
$i^*\sub{SP}$: critical cluster size evaluated by the SP model,
$i^*\sub{CNT}$: critical cluster size evaluated by the CNT (or MCNT)
model, and $\alpha$: sticking probability.}
\begin{center}
\baselineskip 0.1cm
\renewcommand\arraystretch{0.8}
\begin{tabular}{c c c  c l c c c c c c c} \hline    \hline
   run$\#$    & $T $& $T\sub{W} $  & S
       &   $ J\sub{MD}$ & 
           $ J\sub{SP}$
       &   $ J\sub{MCNT}$ &           $ J\sub{CNT}$ & $i_{*}$
       &     $i\sub{*,SP}$ &     $i\sub{*,CNT}$ 
       &  $\alpha $ 
  \\ \hline  

 6e   & 375 & 379 &   3.35 & 4.44$\pm 0.82$  & 0.563  & 985 & 58.4
  & 14 & 20 & 10 &   0.328 \\
 6d   & 375 & 380 &  3.37 & 9.74$\pm 1.6$  & 1.13  & 1461 & 92.1
& 14 & 19  & 10 & 
 0.425 \\
 6c   & 375 & 383 &  3.49 & 19.2$\pm 2.3$  & 3.71  & 2811 & 193.1
 & 13 & 17  & 9 &  
 0.479 \\
 6b   & 375 & 390   & 3.41 & 83.7$\pm 19.0$  & 13.7  & 6116 & 490.8
 & 12 & 16  & 8 &  
  0.671  \\
 5e   & 350 & 357 &   4.61 & 3.71$\pm 0.65$  & 0.67  & 822.1 &  38.5 & 
 13  & 15  & 7 & 
 0.407 \\
 5e-2   & 350 & 355   & 4.98 & 7.03$\pm 3.0$  & 1.29  & 1189 & 53.5 & 
11  & 14  & 6 & 
 0.464 \\
 5d   & 350 & 361    & 4.70 & 8.92$\pm 1.5$  & 2.05  & 1530 & 79.5 & 
12 & 14  & 7 &    0.472 \\
 5c   & 350 & 366  & 4.78 & 29.1$\pm 3.9$  & 9.77  & 3684 & 224.7 &
 14 & 13  &  6 &   0.609 \\
 5b   & 350 & 374 &  4.50 & 68.7$\pm 9.6$  & 27.2  & 7015 & 497.0 & 
 13 & 12  & 6 & 
  0.798  \\
  4e   & 325 & 340 &   6.72 & 3.73$\pm 0.8$  & 1.46  & 911 & 38.0 &
  11 & 11  & 5 &  
  0.476 \\
 4e-2   & 325 & 338  & 7.28 & 5.94$\pm 0.3$  & 2.32  & 1099 & 46.7 & 
  12 & 10  & 5 &   0.649 \\
 4d   & 325 & 346 &  6.54 & 8.63$\pm 0.5$  & 4.79  & 1837 & 88.6  & 
 12 &11  & 5  &
 0.613\\
 4c   & 325 & 352   & 6.48 & 19.1$\pm 4.3$  & 16.8  & 3805 & 213 & 
 11 & 10  & 4 & 
 0.719  \\
 4b   & 325 & 359   & 6.08 & 49.1$\pm 9.8$  & 39.6  & 6630  & 422 &
 11 & 10  & 4 & 
 0.904 \\
 3e   & 300 & 328 &   10.0 & 2.99$\pm 0.4$  & 3.7  & 1076 &  43.9 & 
 10 &9 & 4 & 
 0.622 \\
 3e-2   & 300  & 323 &   11.8 & 4.92$\pm 0.5$  & 7.08  & 1347 & 57.3 & 
 10 & 8  & 3 & 
 0.762 \\
 3d   & 300 & 334 &   9.09 & 7.42$\pm 0.3$  & 8.14  & 1845 &  84.7& 
 11 & 9  &  4 &
  0.794 \\
 3c   & 300 & 340    & 8.73 & 18.3$\pm 0.57$  & 21.5  & 3369 & 175.8 &
 9 & 8  & 3 & 
  0.776 \\
 3b   & 300 & 347   & 7.99 & 46.6$\pm 21$  & 47.8  & 5813 &  343 & 
 10 & 8  &  3 & 
 0.938 \\
 2e   & 275 & 316 &    13.8 & 4.20$\pm 0.60$  & 4.69  & 921.8 & 36.4 &
 9 & 7  &  3 & 
 0.748 \\
 2e-2   & 275 & 310 &   18.6 & 4.50$\pm 0.60$  & 12.2  & 1254 &  53.7 
&  10 & 6  & 2 & 
 1.02 \\
 2d   & 275 & 322 &   13.4 & 5.59$\pm 1.1$  & 13.5  & 1766 & 79.6 & 
 11 & 7  & 3 &  
 0.803 \\
 2c   & 275 & 327 &   12.7 & 16.1$\pm 4.7$  & 33.2  & 3144 & 160   &
10 & 7  & 3  &
 0.911 \\
 2b   & 275 & 335 &   11.2 & 20.0$\pm 4.3$  & 62.4  & 5113 & 287 & 
11 & 7  & 3  &
 1.02\\
 1e   & 250 & 303   & 23.0 & 4.39$\pm 2.0$  & 9.47  & 908 & 37.2 
& 9 & 6  &  2&  
 0.885 \\
 1d   & 250 & 310    & 21.0 & 5.22$\pm 0.95$  & 2.19  & 1623 & 74.1
& 10 & 6  & 2 & 
  0.927 \\
 1c   & 250 & 316    & 18.8 & 10.0$\pm 1.4$  & 42.1  & 2644 & 132  &
 10 & 6  & 2 & 
 1.01  \\
 1b   & 250 & 323   & 16.3 & 24.0$\pm 3.1$  & 82.6  & 4472 & 246 & 
10 & 6  & 2  &
 1.12
\\
\hline   \hline 
\end{tabular}
\vspace{0.8cm}
\end{center}
\end{table*}
}

\end{document}